# Suspicious ARP Activity Detection and Clustering Based on Autoencoder Neural Networks

Yuwei Sun
Graduate School of Information Science
and Technology
University of Tokyo
Tokyo, Japan
ywsun@g.ecc.u-tokyo.ac.jp

Hideya Ochiai
Graduate School of Information Science
and Technology
University of Tokyo
Tokyo, Japan
ochiai@elab.ic.i.u-tokyo.ac.jp

Hiroshi Esaki
Graduate School of Information Science
and Technology
University of Tokyo
Tokyo, Japan
hiroshi@wide.ad.jp

*Abstract*—The rapidly increasing number of smart devices on the Internet necessitates an efficient inspection system for safeguarding our networks from suspicious activities such as Address Resolution Protocol (ARP) probes. In this research, we analyze sequence data of ARP traffic on LAN based on the numerical count and degree of its packets. Moreover, a dynamic threshold is employed to detect underlying suspicious activities, which are further converted into feature vectors to train an unsupervised autoencoder neural network. Then, we leverage K-means clustering to separate the extracted latent features of suspicious activities from the autoencoder into various patterns. Besides, to evaluate the performance, we collect and adopt a real-world network traffic dataset from five different LANs. At last, we successfully detect suspicious ARP patterns varying in scale, lifespan, and regularity on the LANs.

## I. Introduction

The growing popularity of personal computers, mobile devices, and home internet of things (IoT) devices necessitates the need for robustness of network systems against underlying cyber threats. A malware that intrudes into a LAN through a compromised host and tries to expand into other hosts could cause critical issues of data breaches. Besides, in recent years, with more and more cyber-physical systems connecting to the Internet, cyber threats have extended influence to our physical life. As a result, the development of an efficient network traffic inspection system for monitoring and detecting suspicious network events is of great importance.

The research on network traffic classification usually could be grouped into three categories, which are port-based, payload-based, and statistical method [1]. Unfortunately, these methods have shown limitations when it comes to varying network traffic patterns and enormous data for analysis. For instance, attackers tend to not rely on fixed port numbers, instead, often redirecting the ports. For the payload-based method, due to the rapidly increasing network traffic volume, it would be extremely complicated and time-consuming to inspect every packet on a network. Though the statistical method, delicate feature design of network traffic, has achieved great results due to recent years' advancement in deep neural networks (DNNs), it typically relies on supervised approaches with a labeled dataset. Lots of prior research applied a dataset collected from a simulation environment through venerability assessment. In contrast, we applied a dataset from real-world network observation and aimed to detect and cluster various suspicious activity patterns.

ARP is a communication protocol used for discovering the link-layer address of a host such as the MAC address. When malware tries to intrude a LAN, it will first probe into different hosts to find a vulnerability, adopting various attacking methods to hide from being recognized. As a result, the analysis of ARP traffic concerning both the instant volume and changes of it within a period is essential for identifying various suspicious patterns hidden in the ARP traffic. A dynamic threshold was proposed for detecting an initial timepoint of a suspicious event and extracting the complete traffic pattern. Besides, an unsupervised machine learning method of the autoencoder was adapted to compress extracted patterns into latent representations. Finally, to extract various types of patterns, we applied a progressive K-means clustering approach to classify suspicious ARP traffic into various clusters based on the latent representation.

This paper is organized as follows. Section 2 discusses related works about machine learning-based network traffic classification. Section 3 provides an overview of the proposed scheme, consisting of ARP traffic sequence data collection on LANs, suspicious pattern extraction based on the dynamic threshold, and pattern clustering using an autoencoder and K-means. Section 4 presents experiment settings and evaluation results. Section 5 concludes the paper and shows the future work of this research.

## II. Related Work

In recent years, we have seen broad applicability of machine learning, especially deep neural networks, to network intrusion detection due to the great ability of knowledge acquisition from large sets of data. There has been lots of work focusing on supervised network traffic classification, which relied on a set of labeled training data. Unfortunately, labeled real-world network traffic data is usually difficult to obtain because it necessitates enormous experts' efforts in packet inspection. As a result, most of the currently available network intrusion datasets are collected in simulated environments. On the other hand, an unsupervised method is aimed to divide a set of data into various clusters based on data latent feature distribution, which doesn't require humans for the manual labeling. For instance, Kotani et al. [2] demonstrated a flow-based intrusion detection method using a robust autoencoder to detect outliners, using a real-world traffic set called MAWI. Wang et al. [3] presented Softmax classification based on the improved deep belief network (IDBN-SC) for intrusion detection, which showed great performance with the NSL-KDD dataset. Moreover, Purnawansyah et al. [4] proposed a K-means-based bandwidth pattern clustering for network bandwidth management. Fan et al. [5] presented a combined approach of the support vector machine (SVM) and K-means for application traffic clustering and achieved an accuracy rate of over 0.95.

Suspicious ARP traffic detection is aimed to identify underlying suspicious network activities such as zero-day attacks and malicious broadcast requests for finding vulnerable hosts in a network. For example, Whyte et al. [6] proposed network feature-based anomaly detection for host-wise analysis of ARP traffic. Yasami et al. [7] presented a combined approach of K-means clustering and decision trees for unsupervised classification of normal and abnormal events in ARP traffic. Different from the prior research, we aim to classify suspicious ARP traffic patterns through learning optimized latent representations of these patterns with an autoencoder, and further clustering the extracted representations based on K-means.

III. SUSPICIOUS ARP ACTIVITY DETECTION AND CLUSTERING

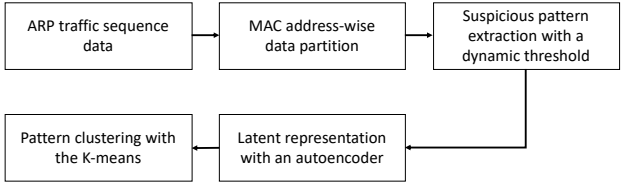

Fig. 1. The scheme of the proposed unsupervised clustering for suspicious ARP traffic data.

We conducted protocol-wise analysis on network traffic sequence data collected from LANs and computed the numbers of total ARP packets sent by each host (with a distinct MAC address) for every five seconds. Then a dynamic threshold computed from the average ARP packet count and degree for each host was adopted to identify initial timepoints of suspicious network events. Then, complete patterns were extracted based on the initial timepoints. Besides, an autoencoder, one kind of unsupervised learning was employed to compress the patterns into latent representations for the K-means-based clustering (Fig. 1).

*A. ARP Traffic Sequence Data Collection on LAN*

A monitoring node deployed on a LAN was applied to capture the broadcast traffic from hosts. Due to our focus is ARP traffic inspection, we first conducted a protocol-wise analysis to extract all ARP traffic sequence data. ARP protocol is used to discover the link layer address of a host such as a MAC address. The number of connected hosts in ARP is also called the degree of destination [8]. Observed abnormal volumes, frequencies, and degrees of outgoing ARP packets usually suggest underlying malicious activities of an adversary. We obtained the sequence data of outgoing ARP traffic and computed the ARP count and degree for every five seconds of each host. The ARP count of a host refers to how many packets are sent by the host to others within five seconds, while the ARP degree refers to how many distinct destination MAC addresses a host sends ARP packets to within five seconds. Besides, other important information such as the existing time of an activity, volume and degree variance with respect to the time, and so on, are considered to contribute to pattern clustering.

*B. Suspicious Pattern Extraction with a Dynamic Threshold*

To extract suspicious patterns from a host's network traffic, we measured time-related changes of the ARP count times the ARP degree. A dynamic threshold drawn from each host's measurement result is employed to detect the initial timepoints of suspicious ARP events, as defined in (1).

$$Threshold = max(128, \frac{\sum_{t=1}^{N} C_t \cdot D_t}{N}) \qquad (1)$$

Where $C_t$ represents the record of ARP count at timepoint *t*, $D_t$ represents the record of ARP degree at timepoint *t*, and *N* is the number of total records. We employed the maximum between a fixed boundary of 128 and a flexible boundary of each host to decide the dynamic threshold.

As such, we applied the dynamic threshold of each host to detect the initial timepoints of suspicious events, tackling the problem of varying daily regular volumes for different hosts (Fig. 2).

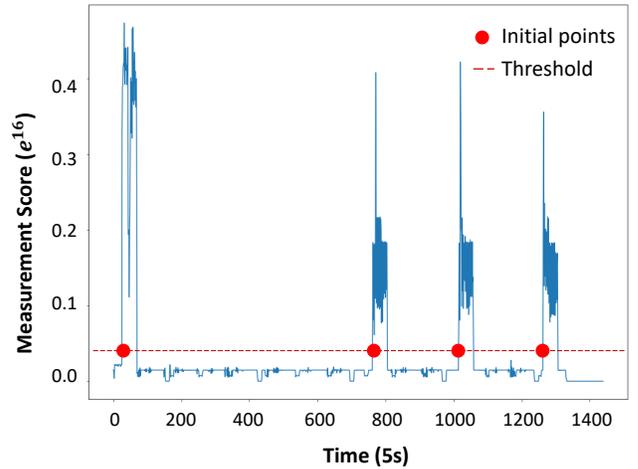

Fig. 2. A sample of suspicious ARP traffic' initial timepoints detection with the dynamic threshold decided by the ARP count and degree sequence data.

Furthermore, after detecting these initial points, we extract ARP information of 300 seconds starting from an initial point as an intact suspicious event observed on LAN. As such, various types of suspicious ARP patterns can be extracted using the dynamic threshold method.

*C. Suspicious Pattern Clustering with Autoencoder Neural Networks and K-Means*

From each extracted event, we generate a feature vector for representing the event based on ARP degree sequence data and ARP frequency sequence data which is the count divided by the degree. Due to each event has a duration of 300 seconds, the obtained feature vector of an event covers a total of 120 items, including 60 items of the ARP degree sequence data and 60 items of the computed ARP frequency sequence data (2). Consequently, we represented features of the various suspicious

ARP events with the generated feature vectors.

$$V_i = \left(D_{i1}, D_{i2}, ..., D_{i60}, \frac{C_{i1}}{D_{i1}}, \frac{C_{i2}}{D_{i2}}, ..., \frac{C_{i60}}{D_{i60}}\right) \quad (2)$$

Where $D_{i1}$ represents the ARP degree recorded within the first five seconds of pattern $i$, and $C_{i1}$ represents the ARP count recorded within the first five seconds of pattern $i$.

Our key idea is that if we can cluster latent representations of ARP events, then we can cluster the corresponding events as well, with the same data partition. To achieve this goal, an autoencoder, one type of unsupervised learning method, is employed. It consists of an encoder and a decoder, where the encoder compresses the input to a latent representation, and the decoder tries to reconstruct the original input from the latent representation. By continually updating the encoder and decoder, the model could learn better and better latent representations of the input data. As such, by employing the trained autoencoder, we can compress the feature vectors from a relatively high dimension of 120 to a low dimension

For the architecture of the autoencoder, we employed a two-layer fully-connected neural network as the encoder, which compresses the input feature vectors from 120 dimensions to 50 dimensions, and finally to 3 dimensions (Fig. 3). On the other hand, the decoder consisting of one fully-connected layer reconstructs the input feature vectors from the compressed three-dimension latent representations. Besides, we employed the ReLU as an activation function at the hidden layer and the output layer of the encoder. We employed the Sigmoid as an activation function at the output layer of the decoder to convert the layer output into a scale of (0, 1). The binary cross-entropy is adopted as the loss function to compute the difference between the reconstruction result from the decoder and the original input data thus updating the model parameters. In addition, we applied the *L2* normalization to convert the input feature vectors into a scale of (0, 1).

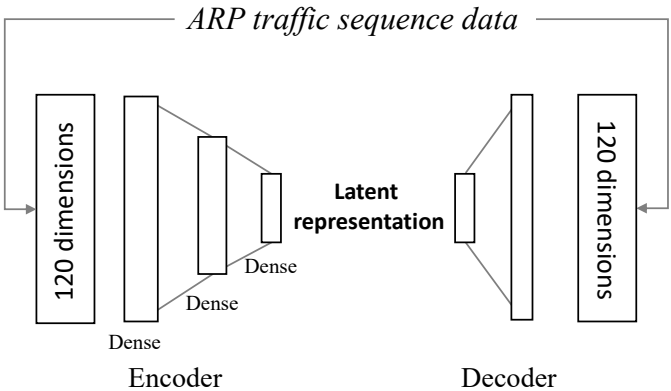

Fig. 3. Architecture of the applied autoencoder to extract the three-dimensional latent representations from input feature vectors.

Then, K-means clustering was used to classify these latent representations based on their similarity with each other using the Euclidean distance. Through finding *k* optimized centroids which partition a set of input into *k* clusters, K-means minimizes the sum of the Euclidean distance between each vector and the centroid of the corresponding cluster defined in (3).

$$D_i = \sum_{i=0}^{k} \sqrt{(S_i - C_i)^2} \quad (3)$$

Where $S_i$ represents data partitioned into the cluster $i$, $C_i$ represents the centroid of the cluster $i$, and $k$ is the total number of partitioned clusters.

Furthermore, to partition data into a suitable number of clusters, we adopted an approach of progressive clustering. The latent representation data were first clustered into a total of five clusters, then an inspection on the clustering result was conducted through clustering visualization. If latent representation data of a cluster covers mixed types of ARP patterns, it will be further partitioned by the K-means method into five new clusters. After the progressive clustering, we can separate mixed multi-type patterns into various clusters.

## IV. EVALUATION

### A. Dataset

We evaluated our model using network traffic data collected from five independent massively distributed LANs in the LAN-Security Monitoring Project [9]. On a LAN, a network traffic monitoring device is connected to a switching hub or a router for data collection. Then the collected data is compressed, encrypted, and transported to the central server for analysis and inspection. Broadcast network traffic and direct traffic sent to the monitoring device are recorded.

For the experiment, we utilized a total of one month's network traffic data between January 1st, 2020 and January 31st, 2020 from five suspicious LANs based on daily observation, these LANs showing instant boosts of traffic volumes and a high variance in traffic sequence data. Based on the aforementioned approaches, a total of 28318 suspicious ARP events with lengths of five minutes were extracted from the dataset, consisting of 6997, 7239, 9498, 41, and 4543 events drawn from each LAN respectively (Table 1). Besides, we computed the average peak traffic volumes of these suspicious patterns, resulting in an average peak ARP count of around 636 per five seconds and an average peak ARP degree of around 290 per five seconds.

TABLE I. DATA SET PROFILE

| LANs | Total MAC Addresses | Suspicious MAC Addresses | Total Extracted Suspicious Patterns |
|---|---|---|---|
| LAN #1 | 108 | 4 | 6997 |
| LAN #2 | 70504 | 839 | 7239 |
| LAN #3 | 1914 | 339 | 9498 |
| LAN #4 | 228 | 10 | 41 |
| LAN #5 | 219 | 61 | 4543 |

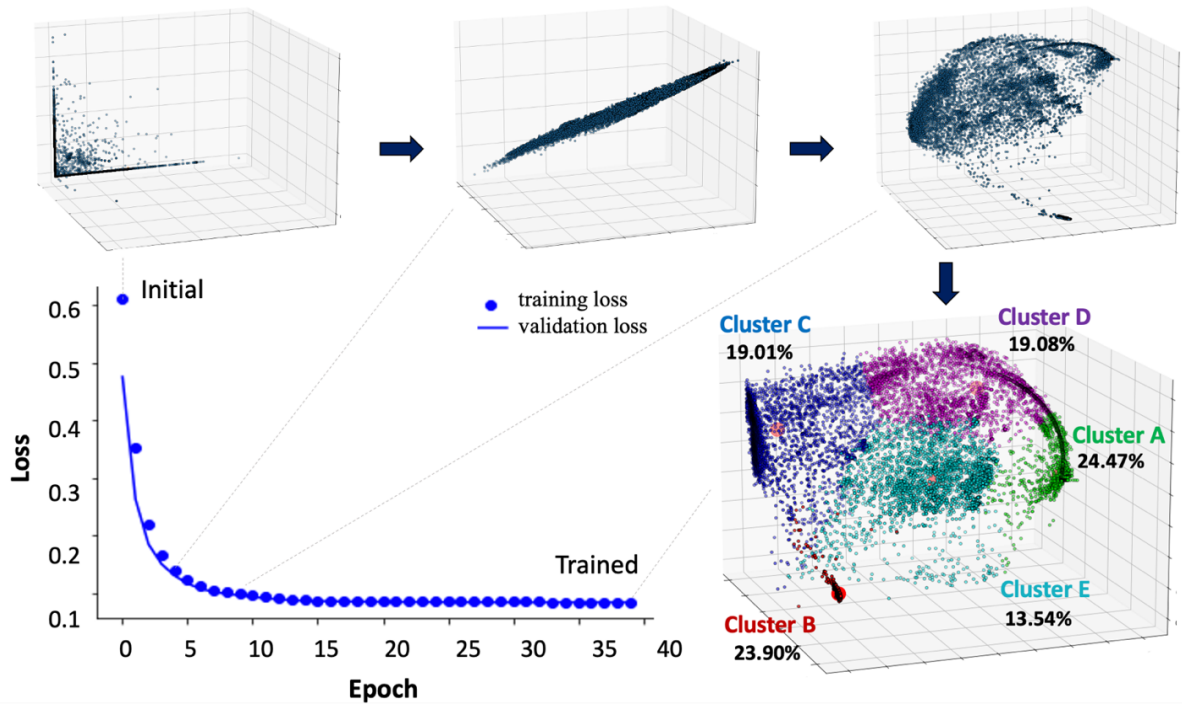

Fig. 4. Changes of patterns' latent representation distribution extracted from the autoencoder during the training progress. At the initial point, traffic patterns tend to gather together. With the progress of learning the latent representation of the input feature vectors, the distributions become sparser and sparser with data converging to various clusters.

*B. Numerical Results*

We trained the autoencoder with a total of 40 epochs and a batch size of 16. We applied as a learning function the Adam using a learning rate of 0.0001 to update model parameters. Besides, the training and validation loss based on the 5-fold cross-validation at each epoch was computed.

As a result, we visualized the distribution of latent representations at different stages of model training. Finally, the progressive clustering with K-means was employed to partition the optimized hidden representations of suspicious ARP events into various clusters (Fig. 4). Besides, for each cluster, we computed the percentage of included data points in all data. Then, based on the Euclidean distance, we further extracted the corresponding ARP events with respect to the latent representation data points closest to the centroids of the partitioned cluster. These extracted patterns are considered the most typical suspicious patterns hidden in ARP traffic (Fig. 5).

By visualizing the ARP event clustering result with K-means, we observed greatly more mixed types of patterns in Cluster C compared with the other clusters. As a result, we further inspected and partitioned data in Cluster C into five new clusters, aiming to separate typical patterns from the mixed data (Fig. 6). The corresponding ARP events with respect to the data points closest to the cluster centroids are shown below (Fig. 7).

Finally, we obtained several typical suspicious ARP events on the LANs. Cluster A shows an instant relatively large boost of traffic. On the other hand, Cluster B shows a repetitively slow probe on the LAN. Cluster D shows an instant relatively small boost of traffic. Cluster E shows a regular series of quick probes in the inspection period. Cluster C1 reveals abnormal continually high traffic. Cluster C2 is similar to Cluster B but less regular. Cluster C4 shows extremely high volumes of the count and degree during a relatively short lifespan with a ratio of 1:1. Cluster C5 shows a repetitive high-volume probe.

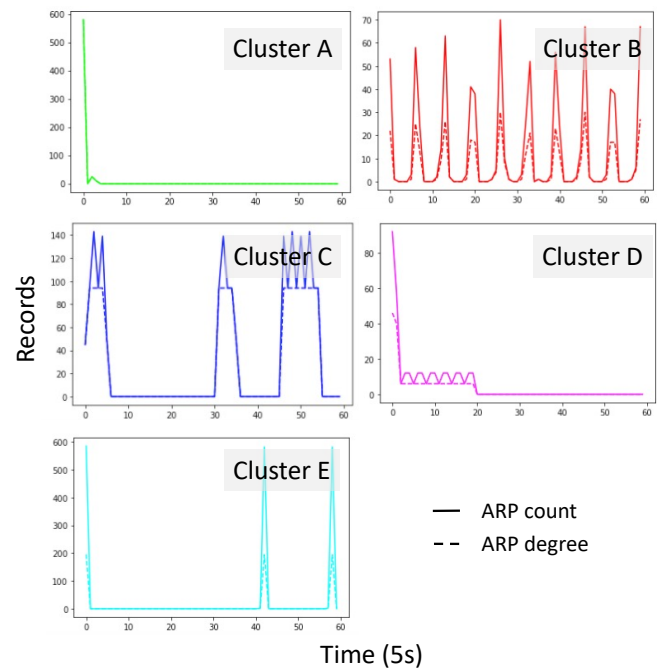

Fig.5. Based on the Euclidean distance, we extracted the corresponding ARP events with respect to the latent representation data points closest to the centroids of the partitioned cluster.

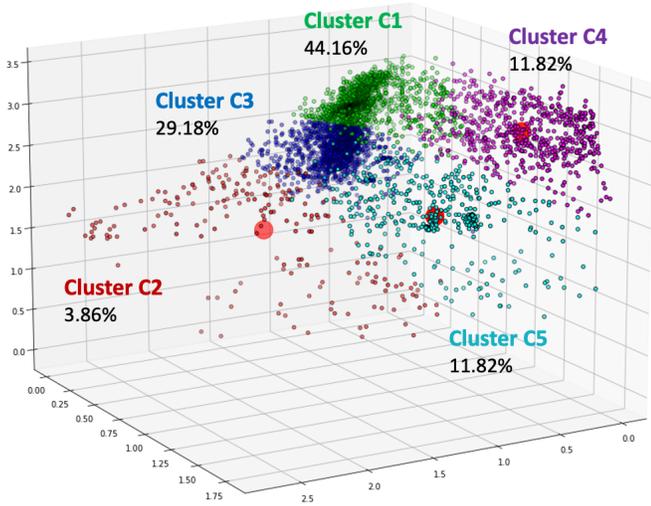

Fig. 6. We further inspected and partitioned data in Cluster C into five new clusters, aiming to separate typical patterns from the mixed data. This graph shows the clustering result and percentages of data in each cluster.

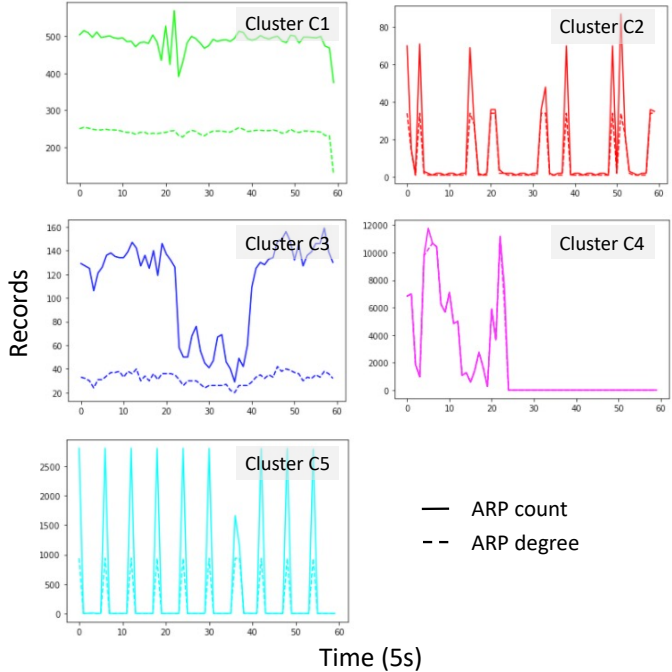

Fig.7. The extracted typical suspicious ARP events from Cluster C.

## V. Conclusion

A dynamic threshold was employed to detect the initial timepoints of underlying suspicious ARP activities on LAN. Then, based on an unsupervised learning method of the autoencoder, we converted feature vectors of these activities into latent representations for dimension reduction. Furthermore, a progressive clustering method using K-means was adopted to separate these latent representations into various patterns. We evaluated the method with a real-world dataset collected from five different LANs. At last, we successfully detect suspicious ARP patterns varying in scale, lifespan, and regularity on the LANs. This work demonstrates an unsupervised method for detecting various patterns of suspicious ARP activities. In the future, a combination with other protocols for unsupervised anomaly detection on LAN is considered.